# Enhanced spectral broadening via self-phase modulation with femtosecond optical pulses in silicon nanowires integrated with 2D graphene oxide films

**Yuning Zhang [1], Jiayang Wu [1,*], Yunyi Yang [1], Yang Qu [1], Linnan Jia [1], Baohua Jia [2,*], and David J. Moss [1,*]**

[1]  Optical Sciences Center, Swinburne University of Technology, Hawthorn, VIC 3122, Australia; yuningzhang@swin.edu.au; yunyiyang@swin.edu.au; yqu@swin.edu.au; ljia@swin.edu.au
[2]  School of Science, RMIT University, Melbourne, VIC 3001, Australia;
\*  Correspondence: jiayangwu@swin.edu.au (J. W.); bjia@rmit.edu.au (B. J.); dmoss@swin.edu.au (D. M.).

**Abstract:** We experimentally demonstrate enhanced spectral broadening of femtosecond optical pulses after propagation through silicon-on-insulator (SOI) nanowire waveguides integrated with two-dimensional (2D) graphene oxide (GO) films. Owing to the strong mode overlap between the SOI nanowires and the GO films with a high Kerr nonlinearity, the self-phase modulation (SPM) process in the hybrid waveguides is significantly enhanced, resulting in greatly improved spectral broadening of the femtosecond optical pulses. A solution-based, transfer-free coating method is used to integrate GO films onto the SOI nanowires with precise control of the film thickness. Detailed SPM measurements using femtosecond optical pulses are carried out, achieving a broadening factor of up to ~4.3 for a device with 0.4-mm-long, 2 layers of GO. By fitting the experimental results with theory, we obtain an improvement in the waveguide nonlinear parameter by a factor of ~3.5 and the effective nonlinear figure of merit (FOM) by a factor of ~3.8, relative to the uncoated waveguide. Finally, we discuss the influence of GO film length on the spectral broadening and compare the nonlinear optical performance of different integrated waveguides coated with GO films. These results confirm the improved nonlinear optical performance for silicon devices integrated with 2D GO films.

## 1. Introduction

As one of the fundamental third-order ($\chi^{(3)}$) nonlinear optical processes, self-phase modulation (SPM) occurs when an optical pulse propagates through a nonlinear medium, where a variation of the medium refractive index caused by the Kerr effect brings about a phase alteration that results in a change in the pulse's spectrum [1-3]. It has been widely studied and used as an applicable all-optical modulation technology for a variety of applications such as broadband optical sources [4, 5], optical spectroscopy [6, 7], optical logic gates [8, 9], pulse compression [10, 11], optical diodes [12, 13], optical modulators / switches [14, 15], and optical coherence tomography [16, 17].

Implementing SPM devices in integrated platforms can provide attractive advantages in achieving compact footprint, high stability, high scalability, and low-cost mass production [18-21]. Although silicon with a high Kerr nonlinearity has been a dominant device platform for integrated photonic devices [22-24], its strong two-photon absorption (TPA) at near-infrared wavelengths results in a poor nonlinear figure-of-merit (FOM = $n_2 / (\lambda \beta_{TPA})$, where $n_2$ is the Kerr nonlinearity, $\beta_{TPA}$ is the two photon absorption coefficient, and $\lambda$ is the wavelength) [25], which poses an intrinsic limitation for the SPM performance in the telecom band. To address this, nonlinear integrated photonic devices have been developed based on other complementary metal-oxide-semiconductor (CMOS) compatible platforms such as silicon nitride ($Si_3N_4$) and high-index doped silica glass (Hydex), which have lower TPA and higher nonlinear FOMs in the telecom band. However, their low Kerr nonlinearity ($n_2 = \sim 2.6 \times 10^{-19}$ m² W⁻¹ and $\sim 1.3 \times 10^{-19}$ m² W⁻¹ for $Si_3N_4$ and Hydex, respectively, over ten times lower than silicon [26, 27]) still limit the nonlinear efficiencies of $Si_3N_4$ and Hydex devices [28, 29].

To overcome the limitations of the existing integrated platforms, the on-chip integration of two-dimensional (2D) materials with ultrahigh Kerr nonlinearity has proven to be a promising approach. Enhanced SPM-induced spectral broadening has been observed in integrated waveguides incorporating graphene [30-32], $MoS_2$ [33], $WS_2$ [34], and graphene oxide (GO) [35]. Among these 2D materials, GO has exhibited a series of distinctive material properties and shows many advantages for implementing hybrid integrated photonic devices with excellent nonlinear optical performance. It has been reported that GO has a large Kerr nonlinearity ($n_2$) that is about 4 orders of magnitude higher than silicon [36, 37] as well as a low linear absorption that is over 100 times lower than graphene at near infrared wavelengths [38, 39]. Moreover, the GO film thickness, length, and position on integrated devices can be precisely controlled by using a facile solution-based, transfer-free coating method and CMOS-compatible film patterning techniques [40, 41]. Previously, we demonstrated enhanced SPM-induced spectral broadening of picosecond optical pulses in SOI nanowires integrated with 2D GO films, achieving a maximum spectral broadening factor (BF) of ~3.8 for a device with 2.2-mm-long, 2 layers of GO [35].

In this paper, we report the experimental observation of significantly enhanced SPM-induced spectral broadening of femtosecond optical pulses after transmission through SOI nanowires integrated with 2D GO films. We fabricate GO-coated SOI nanowires with precise control of the GO film thicknesses and coating length. SPM measurements are performed using femtosecond optical pulses. The GO-coated waveguides show more significant spectral broadening than the uncoated SOI nanowire, achieving a maximum BF of ~4.3 for a device with 2 layers of GO. We also fit the experimental results with theory and obtain improved waveguide nonlinear parameter by up to ~3.5 times and the nonlinear FOM by ~3.8 times. Finally, we discuss the influence of GO film's length on the spectral broadening and compare the nonlinear optical performance of different integrated waveguides incorporating GO films. These results confirm the strong potential of integrating 2D GO films to improve the nonlinear optical performance of silicon photonic devices.

## 2. Device and Characterization

Figure 1a shows a schematic of a GO-coated SOI nanowire waveguide with a monolayer GO film. The bare SOI nanowire has a cross-section of 500 nm × 220 nm, which was fabricated on an SOI wafer with a 220-nm-thick top silicon layer and a 2-μm-thick buried oxide (BOX) layer via CMOS-compatible fabrication processes. First, 248-nm deep ultraviolet photolithography was employed for defining the waveguide layout on photoresist and then the layout was transferred to the top silicon layer by using inductively coupled plasma etching. Next, a 1.5-μm thick silica upper cladding layer was deposited using plasma enhanced chemical vapor deposition (PECVD), followed by opening a window on it down to the BOX layer via photolithography and reactive ion etching processes. Finally, the 2D layered GO film was coated onto the SOI nanowires by using a solution-based method that enabled transfer-free and layer-by-layer film coating, as reported previously [39–42]. Compared to the sophisticated film transfer processes used for other 2D materials like graphene and TMDCs [32, 33, 43], our GO coating method shows advantages in achieving highly scalable fabrication, precise control of the layer number (i.e., film thickness), and conformal film attachment onto integrated devices [37, 41]. Figures 1b-i and b-ii show a schematic cross section and the transverse electric (TE) mode profile of the GO-coated SOI nanowire in Figure 1a, respectively. The interaction between the waveguide evanescent field and the GO film with an ultrahigh Kerr nonlinearity enables enhanced SPM process in the hybrid waveguide and hence improved spectral broadening of optical pulses after propagation through it.

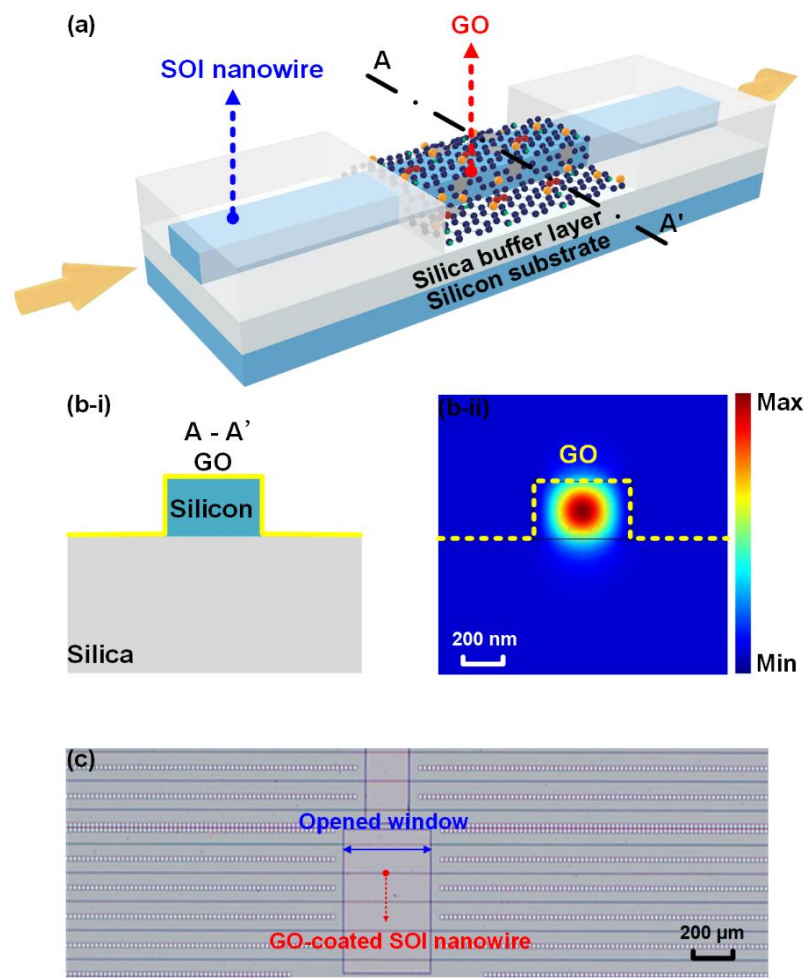

**Figure 1.** (a) Schematic illustration of a GO-coated SOI nanowire with a monolayer GO film. (b-i) Schematic illustration of the cross section and (b-ii) the corresponding TE mode profile of the GO-

coated SOI nanowire in (a). (c) Microscope image of an SOI chip uniformly coated with a monolayer GO film.

Figure 1c presents a microscope image of an SOI chip uniformly coated with a monolayer GO film, showing the good morphology, high uniformity, and high transmittance of the coated GO film. The opened window on the silica upper cladding allows for the control of the GO film's length and position. Based on our previous measurements [35, 38, 39], the GO film thickness shows a near linear relationship at small (< 100) layer numbers, with thickness for each GO layer being ~2.0 nm. For the GO-coated SOI nanowires used in the following SPM measurements, the measured film thicknesses for 1 and 2 layers of GO are ~2.0 nm and ~ 4.2 nm, respectively.

## 3. Loss Measurements

Figure 2 shows the experimental setup employed for both loss and SPM measurements using GO-coated SOI nanowires. Two laser sources were employed, including a continuous-wave (CW) laser and a femtosecond fiber pulsed laser (FPL) that can generate nearly Fourier-transform limited femtosecond optical pulses (pulse duration: ~180 fs, repetition rate: 60 MHz) centered at a wavelength of ~1556.6 nm. After the laser source, an optical isolator was used to prevent the reflected light from damaging it. The power and polarization of the input light were adjusted by a variable optical attenuator (VOA) and a polarization controller (PC), respectively. We injected TE polarized light into the device under test (DUT) for both the loss and SPM measurements because it supports much stronger in-plane interaction between the waveguide evanescent field and the 2D GO film as compared with the out-of-plane interaction [39, 44]. The light was coupled into and out of the DUT by using inverse-taper couplers at both ends of the SOI nanowires, which were butt coupled to lensed fibers with a coupling loss of ~5 dB per facet.

For the loss measurements, two optical power meters, i.e., OPM 1 and OPM 2, were used to measure the power of the light before and after passing the DUT. The losses of the uncoated and GO-coated SOI nanowires were measured by using both the CW laser and the femtosecond FPL. The corresponding results measured by FPL are shown in Figure 3. The total length of the SOI nanowires (including the segments with and without silica cladding) was 3 mm, and the length of the opened window was 0.4 mm. Unless otherwise specified, the powers of CW light and femtosecond optical pulses used for loss and SPM measurements in this paper indicates the powers coupled into the waveguide after excluding the fiber-to-chip coupling loss.

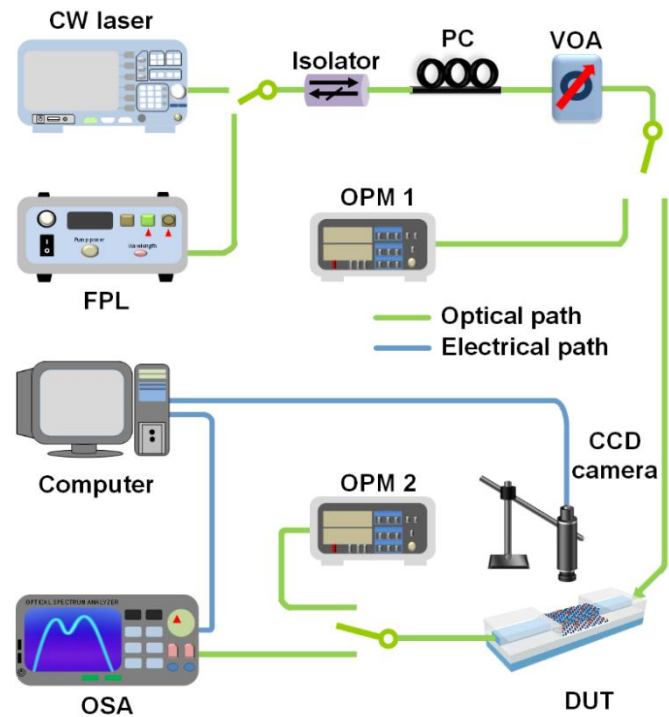

**Figure 2**. Experimental setup for measuring loss and SPM of GO-coated SOI nanowires. CW laser: continuous-wave laser. FPL: fiber pulsed laser. PC: polarization controller. VOA: variable optical attenuator. OPM: optical power meter. DUT: device under test. CCD: charged-coupled device. OSA: optical spectrum analyzer.

We first measured the insertion losses of the uncoated and GO-coated SOI nanowires with 1 and 2 layers of GO using a CW light with a power of ~0 dBm. According to our measurements with the bare SOI nanowire, its propagation loss was ~0.4 dB/mm. The propagation losses of the hybrid waveguides with 1 and 2 layers of GO were ~2.4 dB/mm and ~3.9 dB/mm, corresponding to excess propagation losses of ~2.0 dB/mm and ~3.5 dB/mm induced by the GO films, respectively. These values are higher than those of GO-coated $Si_3N_4$ and Hydex waveguides [38, 40], mainly due to the much stronger GO mode overlap in the GO-coated SOI nanowires. The excess propagation loss induced by GO is about two orders magnitude lower than the excess propagation loss induced by graphene in graphene-coated SOI nanowires [32, 45], reflecting the low loss of GO compare to graphene and its great potential for implementing nonlinear photonic devices requiring high optical powers.

Figure 3a depicts the power-dependent excess insertion loss of the GO-coated and uncoated SOI nanowires (EIL, defined as the additional insertion loss) versus the coupled peak power of femtosecond optical pulses. The input peak power ranges from 98 W to 160 W, corresponding to an average input power range of 1.1 mW – 1.7 mW. By measuring the loss using a CW light with an average power in the same range, we did not observe any obvious power-dependent variations of the insertion losses of the hybrid waveguides, reflecting that the photo-thermal changes in the GO films were negligible in this power range [38, 39]. For the bare SOI nanowire ($N$ = 0), the EIL increases with peak power primarily due to the TPA and free carrier absorption of silicon [46]. For the GO-coated SOI nanowires, the measured EIL is slightly lower than that for the uncoated SOI nanowire, with more obvious difference for the device with 2 layers of GO compared to the device with 1 layer of GO. This reflects that in the hybrid waveguides there exists saturable absorption (SA) induced by the GO films, which has also been observed in our previous work [35]. Figure 3b depicts the SA-induced excess propagation loss ($\Delta SA$) extracted from Figure 3a after excluding the excess propagation loss induced by the bare SOI nanowire. The negative values of $\Delta SA$ reflect that the SA-induced loss decreases with the peak power, showing an opposite trend to TPA where the loss increases with light power [25, 47, 48]. In our measurements, we also note that the change in the loss of the hybrid waveguides was not permanent, and the measured *EIL* in Figure 3a was repeatable.

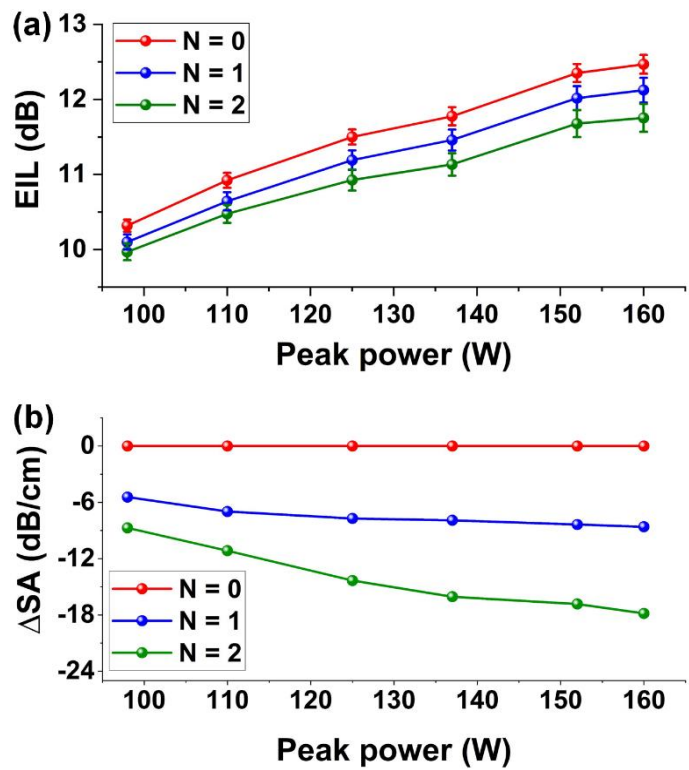

**Figure 3**. (a) Measured excess insertion loss (*EIL*) of GO-coated SOI nanowires versus input power of optical pulses. (b) Excess propagation loss induced by the SA ($\Delta SA$) versus peak power of input optical pulses. In (a) – (b), the results for uncoated ($N = 0$) and hybrid SOI nanowires coated with 1 and 2 layers of GO ($N = 1$, 2) are shown for comparison. The data points depict the average of measurements on three samples and the error bars illustrate the variations among the different samples.

## 4. SPM Measurements

In the SPM measurements, we used the same FPL and the same fabricated devices as those used for loss measurements in Section 3 to measure the SPM-induced spectral broadening of femotosecond optical pulses. As shown in Figure 2, the input optical pulses generated by the FPL were coupled into the DUT, and the output signal was then sent to an optical spectrum analyzer (OSA) to observe the spectral broadening. The corresponding results are shown in Figure 4.

Figure 4a shows the normalized input and output spectra of femotosecond optical pulses after going through the bare and GO-coated SOI nanowires. The peak power of the input picosecond optical pulses was kept the same at ~160 W. Due to the high Kerr nonlinearity of silicon, the output spectrum after passing the uncoated SOI nanowire shows obvious spectral broadening than the input pulse spectrum. For the hybrid device with 1 layer of GO, the output spectrum shows more significant spectral broadening than the bare SOI nanowire, and the device with 2 layers of GO shows more significant spectral broadening than the device with 1 layer of GO. This reflects enhanced SPM induced by the high Kerr nonlinearity of the GO films and the strong GO mode overlap in the GO-coated SOI nanowires.

Figures 4b and 4c show the output spectra after propagation through the hybrid waveguides with 1 and 2 layers of GO measured using femtosecond optical pulses with different peak powers, respectively. We chose 6 different input peak powers ranging from 98 W to 160 W – the same as those in Figure 3. As expected, the spectral broadening of the output spectrum becomes more significant as the peak power increases.

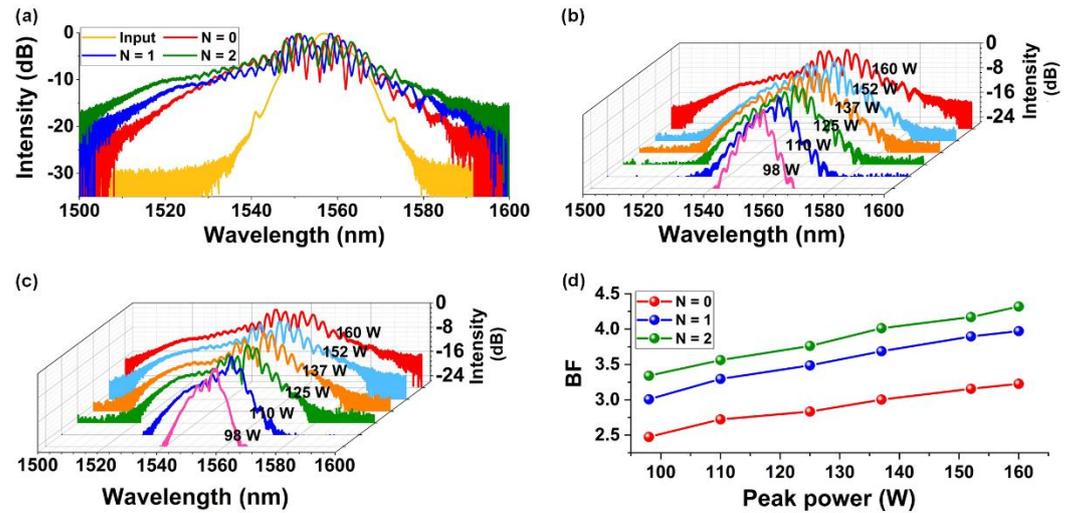

**Figure 4.** SPM experimental results. (a) Normalized spectra of optical pulses before and after propagation through the GO-coated SOI nanowires with 1 and 2 layers of GO at an input peak power of ~160 W. (b) – (c) Optical spectra measured at different input peak powers for the hybrid waveguides with 1 and 2 layers of GO, respectively. (d) BFs of the measured output spectra versus input peak power for the hybrid waveguides with 1 and 2 layers of GO. In (a) – (d), the corresponding results for the uncoated SOI nanowires ($N = 0$) are also shown for comparison.

To quantitatively analyze and compare the spectral broadening of femtosecond optical pulses in these waveguides, we calculated the BFs for the measured output spectra defined as [31, 35, 49]:

$$BF = \frac{\Delta\omega_{rms}}{\Delta\omega_0} \tag{1}$$

where $\Delta\omega_0$ and $\Delta\omega_{rms}$ are the root-mean-square (RMS) spectral widths of the input and output signals, respectively.

Figure 4d shows the BF for the GO-coated SOI nanowires as a function of the peak power of femtosecond optical pulses. The results for the bare SOI nanowire are also shown for comparison. As can be seen, the BF for the hybrid waveguides is higher than that of the bare waveguide, and the BF for the hybrid waveguide integrated with 1 layer of GO is lower than that for the device with 2 layers of GO, showing agreement with the results in Figure 4a. The BF increases with the input peak power, which is consistent with the results in Figures 4b and 4c. At the peak power of ~160 W, a maximum BF of ~4.3 is achieved for the hybrid waveguide with 2 layers of GO.

## 5. Theoretically Analysis and Discussion

According to the theory in Refs. [32, 35, 46], we modeled the SPM process in the GO-coated SOI nanowires as follows:

$$\frac{\partial A}{\partial z} = -\frac{i\beta_2}{2}\frac{\partial^2 A}{\partial t^2} + i\gamma \, |A|^2 A - \frac{1}{2}i\sigma\mu N_c A - \frac{1}{2}\alpha \, A \qquad (2)$$

where $i = \sqrt{1}$, $A(z, t)$ is the slowly varying temporal pulse envelope along the propagation direction $z$ of the waveguide, $\beta_2$ is the second-order dispersion coefficient, $\gamma$ is the waveguide nonlinear parameter, $\sigma$, $\mu$, and $N_c$ are the FCA coefficient, free carrier dispersion (FCD) coefficient, and free carrier density in silicon, respectively, and $\alpha$ is the total loss including both linear loss and nonlinear loss. The nonlinear loss includes TPA and FCA losses of bare SOI nanowires and SA loss induced by the GO films. In Equation (2), we keep only the $\beta_2$ item since the physical length of the waveguides is smaller than the dispersion length [50].

Based on Equation (2), we fit the measured spectra to obtain the nonlinear parameters ($\gamma$'s) for the bare and hybrid waveguides. The GO-coated SOI nanowires were divided into uncoated (with silica cladding) and hybrid segments (coated with GO films) to perform numerical calculation, where the output from the previous segment was set as the input for the subsequent one. We obtain a fit $\gamma$ of ~288 W$^{-1}$·m$^{-1}$ for the bare SOI nanowire and fit $\gamma$'s for the hybrid waveguides with 1 and 2 layers of GO of ~675 W$^{-1}$·m$^{-1}$ and ~998 W$^{-1}$·m$^{-1}$, respectively, which are ~2.3 and ~3.5 times that of the bare SOI nanowire. These results show a good agreement with the previous work [35], indicating the high consistency and further confirming the remarkably improved Kerr nonlinearity for the hybrid waveguides.

Based on the fit $\gamma$'s of the hybrid waveguides, we further extract the Kerr coefficient ($n_2$) of the layered GO films using [40, 51, 52] :

$$\gamma = \frac{2\pi}{\lambda_c} \frac{\iint_D \, n_0^2(x, y) n_2(x, y) S_z^2 dx dy}{\left[\iint_D \, n_0(x, y) S_z dx dy\right]^2} \qquad (3)$$

where $\lambda_c$ is the pulse central wavelength, $D$ is the integral of the optical fields over the material regions, $S_z$ is the time-averaged Poynting vector calculated using mode solving software, $n_0(x, y)$ and $n_2(x, y)$ are the refractive index profiles calculated over the waveguide cross section and the Kerr coefficient of the different material regions, respectively. The values of $n_2$ for silica and silicon used in our calculation were $2.60 \times 10^{-20}$ m$^2$ W$^{-1}$ [26] and $6.0 \times 10^{-18}$ m$^2$ W$^{-1}$, respectively, with the latter obtained by fitting the experimental results for the bare SOI nanowire.

The extracted $n_2$ of 1 and 2 layers of GO are ~$1.45 \times 10^{-14}$ m$^2$ W$^{-1}$ and ~$1.36 \times 10^{-14}$ m$^2$ W$^{-1}$, respectively, which are about 4 orders of magnitude higher than that of silicon and agree reasonably well with our previous measurements [35]. Note that $n_2$ of 2 layers of GO is lower than that of 1 layer of GO, which we infer it may arise from the increased inhomogeneous defects within the GO layers and imperfect contact between the multiple GO layers.

Previously, we have fabricated SOI nanowires with much thicker GO films (up to 20 layers [35]), and have performed detailed theoretical analysis for the influence of GO film's length, thickness, and coating position on the nonlinear performance of GO-coated integrated waveguides [28, 29]. In this work, the maximum GO film thickness for the fabricated device is 2 layers (i.e., ~4.2 nm), which is mainly used to compare the spectral broadening performance for the picosecond and femtosecond optical pulses.

Based on the SPM modeling in Equation (2) and the fit parameters of GO, we compare the BFs of femtosecond and picosecond optical pulses after transmission through the same GO-coated SOI nanowires. Figures 5a and 5b shows the BFs of femtosecond and picosecond optical pulses versus GO film length ($L_c$) for the hybrid waveguides with 1 and 2 layers of GO, respectively. The corresponding results for the bare SOI nanowire with constant BFs are also shown for comparison. Both the picosecond and femtosecond optical pulses had the same repetition rate of ~60 MHz. For the picosecond pulses with a pulse duration of ~3.9 ps, the average input power is 3 mW, corresponding to a peak power range of 13 W – the same as that used in our previous experiment [35]. For the femtosecond pulses with a pulse duration of ~180 fs, the average input power is 1.7 mW, corresponding to a peak power of 160 W – the same as that used for the SPM measurements in Section 4.

In Figure 5a, the BFs of femtosecond optical pulses after propagation through the hybrid waveguides first increase with $L_c$ and then decrease, with the maximum values being achieved at intermediate film lengths. The optimized film length corresponding to the maximum BF for the device with 2 layers of GO is smaller than that for the device with 1 layer of GO. This reflects that the Kerr nonlinearity enhancement dominates for the devices with relatively small $L_c$ and layer number $N$, and the influence of loss increase becomes more significant as $L_c$ and $N$ increase.

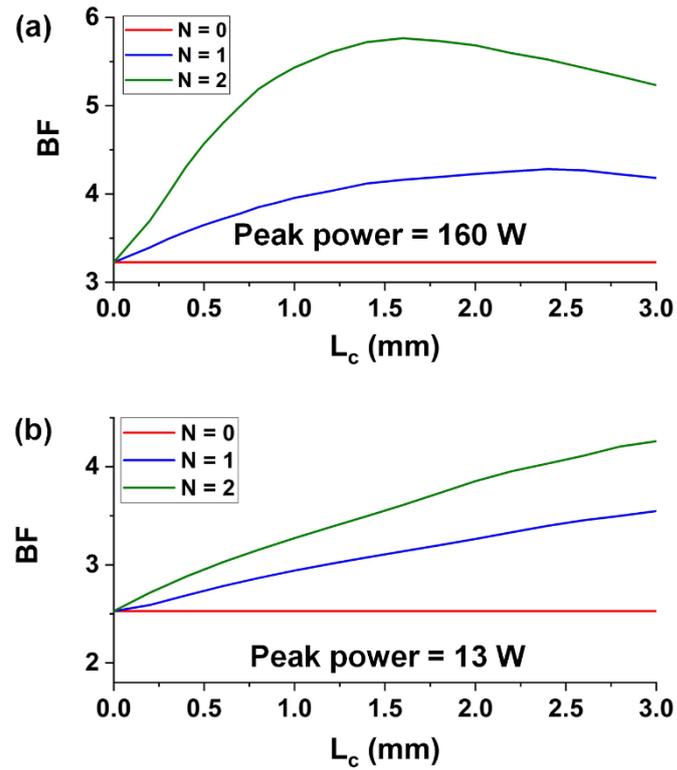

**Figure 5.** (a) BFs of femtosecond optical pulses versus GO film length ($L_c$) for the hybrid waveguides with 1 and 2 layers of GO. The peak power is ~160 W. (b) BFs of picosecond optical pulses versus GO film length ($L_c$) for the hybrid waveguides with 1 and 2 layers of GO. The peak power is ~13 W. In (a) and (b), the corresponding results for uncoated waveguides ($N = 0$) are also shown for comparison.

For the picosecond optical pulses in Figure 5b, the BFs after propagation through the hybrid waveguides are lower than the BFs of the femtosecond optical pulses at the same $L_c$. This is because the femtosecond optical pulses with a much higher peak power drive more significant SPM in the hybrid waveguides. Unlike the trend in Figure 5a, the BFs in Figure 5b increase monotonically with $L_c$. This is mainly due to the fact that the higher peak power of femtosecond optical pulses also induces higher TPA of the SOI nanowires, resulting in a trade-off between both enhanced SPM and TPA in the hybrid waveguides.

In Table 1, we compare the nonlinear performance of different integrated waveguides incorporating GO, along with corresponding results for the bare waveguides. As can be seen, the fit $\gamma$, $n_2$, $FOM_1$ and $FOM_2$ of GO-coated SOI nanowires in this work show good agreement with those in the previous work obtained by fitting the experimental results of picosecond optical pulses [35], highlighting the high consistency of our GO films. We calculated two different figure-of-merits, i.e., $FOM_1$ and $FOM_2$, which are widely studied and exploited in comparing nonlinear optical performance. The former one is defined from the perspective of nonlinear absorption [25, 26], whereas the latter one is defined based on the trade-off between Kerr nonlinearity and linear loss [53]. Interestingly, the two FOMs show contrary results for the different hybrid waveguides. The $FOM_1$ of the GO-coated SOI nanowires is lower than the other two waveguides mainly due to the strong

TPA of silicon, whereas its $FOM_2$ is much higher resulting from the large $n_2$ of silicon and its strong GO mode overlap.

**Table 1.** Comparison of nonlinear optical performance of different integrated waveguides incorporating GO. FOM: figure of merit.

| Integrated waveguide | GO layer number [a] | Waveguide dimension (μm) | $\gamma$ (W⁻¹·m⁻¹) [b] | Fit $n_2$ (×10⁻¹⁴ m²/W) | $PL$ (dB/cm) [c] | $FOM_1$ (a. u.) [d] | $FOM_2$ (W⁻¹) [e] | Ref. |
|---|---|---|---|---|---|---|---|---|
| SOI | $N = 0$ | 0.50 × 0.22 | 288.00 | 6.00 × 10⁻⁴ | 4.30 | 0.74 | 0.75 | [35] |
| | $N = 1$ | | 668.01 | 1.42 | 24.80 | 2.07 | 0.96 | |
| | $N = 2$ | | 990.23 | 1.33 | 38.91 | 2.81 | 1.03 | |
| SOI | $N = 0$ | 0.50 × 0.22 | 288.00 | 6.00 × 10⁻⁴ | 4.30 | 0.74 | 0.75 | This work |
| | $N = 1$ | | 675.15 | 1.45 | 24.60 | 2.08 | 0.97 | |
| | $N = 2$ | | 998.18 | 1.36 | 38.52 | 2.83 | 1.05 | |
| Si₃N₄ | $N = 0$ | 1.60 × 0.66 | 1.51 | 2.60 × 10⁻⁵ | 3.00 | | 0.016 | [38] |
| | $N = 1$ | | 13.14 | 1.41 | 6.05 | ≫ 1 | 0.089 | |
| | $N = 2$ | | 28.23 | 1.35 | 12.25 | | 0.099 | |
| Hydex | $N = 0$ | 2.00 × 1.50 | 0.28 | 1.28 × 10⁻⁵ | 0.24 | | 0.004 | [40] |
| | $N = 1$ | | — | — | 1.26 | ≫ 1 | 0.007 | |
| | $N = 2$ | | 0.90 | 1.5 | 2.23 | | 0.009 | |

[a] $N = 0$ corresponds to the results for the uncoated SOI nanowire, Si₃N₄, and Hydex waveguides, whereas $N = 1$ and 2 correspond to the results for the hybrid waveguides with 1 and 2 layers of GO, respectively.

[b] $\gamma$ is the nonlinear parameter. For the hybrid waveguides, $\gamma$'s are the effective values calculated based on Refs. [38, 40].

[c] $PL$ is the linear propagation loss of the GO-coated waveguides.

[d] The definition of $FOM_1 = n_2 / (\lambda \beta_{TPA})$ is the same as those in Refs. [25, 26], with $n_2$ and $\beta_{TPA}$ denoting the effective Kerr coefficient and TPA coefficient of the waveguides, respectively, and $\lambda$ denoting the light wavelength. The values for the Si₃N₄, and Hydex waveguides are ≫ 1 due to negligible TPA observed in these waveguides.

[e] The definition of $FOM_2 = \gamma \times L_{eff}$ is the same as that in Ref. [53]. Here, the GO films are uniformly integrated on the waveguides and the waveguides length for the SOI nanowire, Si₃N₄, and Hydex waveguides are 3 mm, 20 mm, 15 mm.

In Table 1, the $FOM_2$ is a function of waveguide length $L$ given by [53]

$$FOM_2 (L) = \gamma \times L_{eff} (L) \qquad (4)$$

where $L_{eff} (L) = [1 - exp (-\alpha_L \times L)] / \alpha_L$ is the effective interaction length, with $\gamma$ and $\alpha_L$ denoting the waveguide nonlinear parameter and the linear loss attenuation coefficient, respectively. Figure 6 shows $L_{eff}$ and $FOM_2$ versus waveguide length ($L$) for SOI nanowires uniformly coated with 1 and 2 GO layers, together with the result for the bare waveguide (i.e., $N = 0$). $FOM_2$ first rapidly increases with $L$ and then grows more progressively as $L$ becomes longer. For a shorter $L$, the $FOM_2$'s of the hybrid waveguides are higher than that of comparable bare waveguide, whereas the $FOM_2$ of the bare waveguide gradually approaches and even exceeds those of the hybrid waveguides when $L$ increases. This reflects that the negative influence induced by increased loss becomes more dominant as $L$ increases.

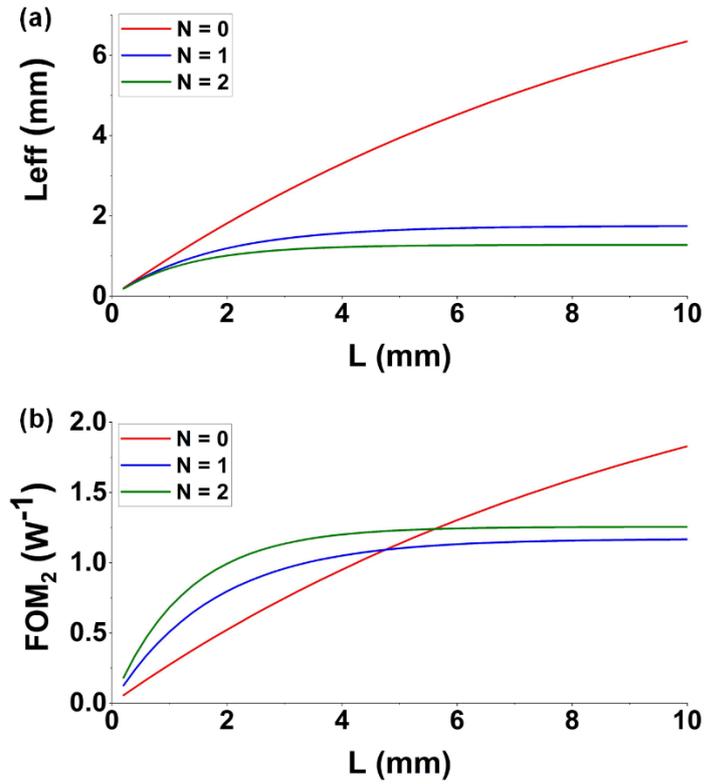

**Figure 6.** (a) Effective interaction length ($L_{eff}$) versus waveguide length ($L$) for GO-coated SOI nanowires uniformly coated with 1 and 2 layers of GO. (b) FOM₂ versus waveguide length ($L$) for hybrid waveguides uniformly coated with 1 and 2 layers of GO. In (a) and (b), the corresponding results for uncoated waveguides ($N = 0$) are also shown for comparison.

## 6. Conclusions

We demonstrate an enhanced SPM-induced spectral broadening of femtosecond optical pulses after propagation through SOI nanowires integrated with 2D GO films. By using a solution-based, transfer-free coating method, we achieve the integration of GO films onto SOI nanowires with precise control of the film thickness. We perform detailed SPM measurements using the fabricated devices, achieving a maximum BF of ~4.3 for a device with 2 layers of GO. The experimental results agree well with theory, showing improved nonlinear parameter of up to 3.5 times and nonlinear FOM of up to 3.8 times compared to the bare waveguide. Finally, the influence of GO film's length on the spectral broadening is analyzed and the nonlinear optical performance of different GO-coated integrated waveguides is compared. This work verifies the effectiveness of improving the nonlinear performance for silicon photonic devices through integration of 2D GO films.

**Author Contributions:** Conceptualization, methodology, Y.Z., J.W.; software, Y.Z., Y.Q., J.W.; silicon device design and fabrication, J.W., Y.Y.; GO film coating and material characterization, Y.Y., L.

J.; writing—original draft preparation, Y.Z.; writing—review and editing, Y.Z., J.W., D.M.; supervision, J.W., B.J., D.M..

**Funding:** Please add: This research received no external funding.

**Data Availability Statement:** Not applicable.

**Acknowledgments:** This work was supported by the Australian Research Council Discovery Projects Programs (No. DP150102972 and DP190103186), the Swinburne ECR-SUPRA program, the Australian Research Council Industrial Transformation Training Centers scheme (Grant No. IC180100005), and the Beijing Natural Science Foundation (No. Z180007).

**Conflicts of Interest:** The authors declare no conflict of interest.